\documentclass[letter]{aa} 
\usepackage{graphicx}
\usepackage{txfonts}
\begin{document}

   \title{One more neighbor: The first brown dwarf in the VVV survey\thanks{Based on observations  taken within the ESO VISTA
    Public Survey VVV, Programme ID 179.B-2002}}

   \author{J.~C.~Beam\'in\inst{1}
          \and D.~Minniti\inst{1,2,3}
          \and M.~Gromadzki\inst{4}
          \and R.~Kurtev\inst{4}
          \and V.~D.~Ivanov\inst{5}
          \and Y.~Beletsky\inst{6}
          \and P.~Lucas\inst{7}
          \and R.~K.~Saito\inst{1,2,4,8}
          \and J.~Borissova\inst{4}
          }

   \institute{Instituto de Astrof\'isica, Facultad de F\'isica, Pontificia Universidad Cat\'olica de Chile,
              Casilla 306, Santiago 22, Chile\\
              \email{jcbeamin@astro.puc.cl}
              \and
            The Milky Way Millennium Nucleus, Av. Vicuña Mackenna 4860, 782-0436 Macul, Santiago, Chile
             \and
	      Vatican Observatory, Vatican City State V-00120, Italy 
             \and
             Departamento de Física y Astronomía, Facultad de Ciencias, Universidad de Valparaíso, Ave. Gran Bretaña 1111, Playa Ancha,Valpara\'iso Chile
             \and
             European Southern Observatory, Ave. Alonso de Cordoba 3107, Casilla 19001, Santiago, Chile 
             \and
             Las Campanas Observatory, Carnegie Institution of Washington, Colina el Pino, Casilla 601 La Serena, Chile
             \and
             Centre for Astrophysics Research, Science and Technology Research Inst., University of Hertfordshire, Hatfield AL10 9AB, UK
             \and
              Universidade Federal de Sergipe, Departamento de F\'isica, Av. Marechal Rondon s/n, 49100-000, S\~ao Crist\'ov\~ao, SE, Brazil
             }

   \date{Received July 04, 2013; accepted August 12, 2013}

  \abstract
   { The discovery of brown dwarfs (BDs) in the solar neighborhood and young star clusters 
   has helped to constraint the low-mass end of the stellar mass function and the initial 
   mass function. We use data of the Vista Variables in the V\'ia L\'actea (VVV), a 
   near-infrared (NIR) multi-wavelength ($ZYJH$ $K_{\rm s}$) multi-epoch ($K_{\rm s}$) 
   ESO Public Survey mapping the Milky Way bulge and southern Galactic plane to search 
   for nearby BDs. 
   }
   {The ultimate aim of the project is to improve the completeness of the census of
   nearby stellar and substellar objects towards the Galactic bulge and inner disk regions.  
   }
   { Taking advantage of the homogeneous sample of VVV multi-epoch data, we identified
   stars with high proper motion ($\geq$ 0.1\arcsec yr$^{-1}$), and then selected low-mass objects 
   using NIR colors. We searched for a possible parallax signature using the all available $K_{\rm s}$ 
   band epochs. We set some constraints on the month-to-year scale
   $K_{\rm s}$ band variability of our candidates, and even searched for
   possible transiting companions. 
   We obtained NIR spectra to properly classify spectral type and then the physical
   properties of the final list of candidates.
   }
   {We report the discovery of \object{VVV BD001}, a new member of the local volume-limited sample
   (within 20\,pc from the sun) with well defined proper motion, distance, and luminosity.
   The spectral type of this new object is an L5$\pm$1, unusually blue dwarf. 
   The proper motion for this BD is PM($\alpha$)=-0.5455$\pm$0.004 \arcsec  yr$^{-1}$, PM($\delta$)=-0.3255$\pm$0.004 \arcsec  yr$^{-1}$,
    and it has a parallax of 57$\pm$4 mas which translates into a distance of 17.5 $\pm$ 1.1 pc. 
    \object{VVV BD001} shows no evidence of variability ($\Delta K_{\rm s}$  <0.05mag) over two years, 
   especially constrained on a six month scale during the year 2012. 
   }
   {}
   
\authorrunning{Beam\'in et al.}  
\titlerunning{Discovery of the first brown dwarf in the VVV survey}
\maketitle
%

\section{Introduction}\label{Intro}
Brown dwarfs (BD) are substellar objects with very low surface temperatures
(300$\lesssim$T$\lesssim$2200K) that are unable to sustain hydrogen fusion
in their interiors.
The presence of molecules in their atmospheres, such H$_2$O, CO, and CH$_4$,
is important at  these low temperatures because the strong molecular
absorption determines their near infrared (NIR) colors \citep{2005nlds.book.....R}.
Different teams working in large surveys have found hundreds of brown dwarfs all over 
the sky using appropriate color constraints (e.g., DENIS , 
2MASS, UKIDSS, and the recent WISE;  \citet{1999A&A...349..236}, \citet{2006AJ....131.1163S}, \citet{2007MNRAS.379.1599L}, 
\citet{2010AJ....140.1868W}). Some of these surveys (i.e., WISE) were optimized to
detect objects with typical colors of BDs.
The only regions in the sky where the color selection might not be optimal are 
very crowded regions, such as the Galactic bulge and the inner Galactic disk where the source confusion,
 and the high level of extinction, make the color selection far less useful than in less dense environments. 

One successful strategy for discovering cool nearby objects is to search high proper motion (HPM) objects using optical
and NIR, or only NIR observations. Various authors have found several ultra-cool dwarfs in 
high stellar density and low galactic latitude regions cross-matching sources from 
DENIS, 2MASS and UKIDSS \citep[e.g.,][]{2007MNRAS.378..901F, 2008MNRAS.383..831P, 2009MNRAS.394..857D, 2010ApJ...718L..38A}.
 The strength of this approach is illustrated by \citet{2013ApJ...767L...1L}, 
 who found the closest BD to the Sun only 2 pc away, 
 using astrometric information from WISE catalogs to select HPM
 objects. This is a binary system composed of an L8 and a T1 BD
\citep{2013arXiv1303.7283B,2013ApJ...770..124K}.

The ESO public survey Vista Variables in the V\'ia L\'actea (VVV)
\citep{2010NewA...15..433M, 2012A&A...537A.107S} has the potential of being a large and 
homogeneous multi-epoch NIR database with high sensitivity and high spatial 
resolution for detecting low-mass stellar and substellar objects via selection 
of HPM sources. The survey is a multiwavelength ($ZYJH$ $K_{\rm s}$ ) photometric survey  observing
towards the Milky Way bulge and southern Galactic plane. Upon completion, the VVV Survey will have a
variability campaign covering a time baseline of around seven years, with about 100 visits 
per pointing in the $K_{\rm s}$  band. The VVV limiting magnitude in the $K_{\rm s}$  band is $\sim$ 4 mag 
deeper than previous large scale near-IR surveys such as 2MASS and DENIS and the spatial
resolution is also better.
The variability campaign will allow us to 1) detect possible eclipses/transits in our HPMs
sources and 2) put some upper limits on the variability of BDs in the $K_{\rm s}$  band. 
Variability of BD is an interesting because it allows constraining cloud 
formation, climate evolution, and magnetic activity that has been proven to exist
in these very low-mass objects \citep[e.g.,][]{2013ApJ...768..121A, 2013ApJ...767..173H}.   
As an example, the cooler component of Luhman's BD binary is highly variable, and it shows changes in the cloud 
structure on a time scale of days \citep{2013arXiv1304.0481G}. The VVV survey will probe BDs variability
on multiple time scales from days up to years.

In Section \ref{obs} we give the observations and discovery method of \object{VVV BD001},
 and in Section \ref{analysis} we analyze the characteristics of the new BD. Finally 
the conclusions and future perspectives are detailed in Section \ref{Conclusions}.

\section{Observations and method}\label{obs}

The VVV survey observations were carried out with the 4.1m VISTA telescope
 at ESO Paranal Observatory, using the VIRCAM instrument\footnote{http://www.eso.org/sci/facilities/paranal/instruments/vircam/}.
This instrument is made up of 16 detectors of 2048x2048 pixels, with a resolution of 
0.34\arcsec/pix.
Each pointing, covering 0.6 deg$^2$, is called a `pawprint', and six overlapping 
pawprints are used to build one final image (Tile) covering twice an area of 
1.5 deg$^2$.
The total area coverage of VVV survey is $\sim$ 560 deg$^2$
The individual pawprints and tiles are processed by the Cambridge Astronomical Unit (CASU). 
They provide aperture photometry and astrometry for the
images\footnote{http://casu.ast.cam.ac.uk/surveys-projects/vista}. The VVV data are publicly available 
through the VISTA Science Archive\footnote{http://horus.roe.ac.uk/vsa/} \citep[VSA,][]{2012A&A...548A.119C}.
More technical information about the VVV survey can be found in \citet{2012A&A...537A.107S,2013A&A...552A.101S}

   \begin{figure}
   \centering
   \includegraphics[scale=0.5]{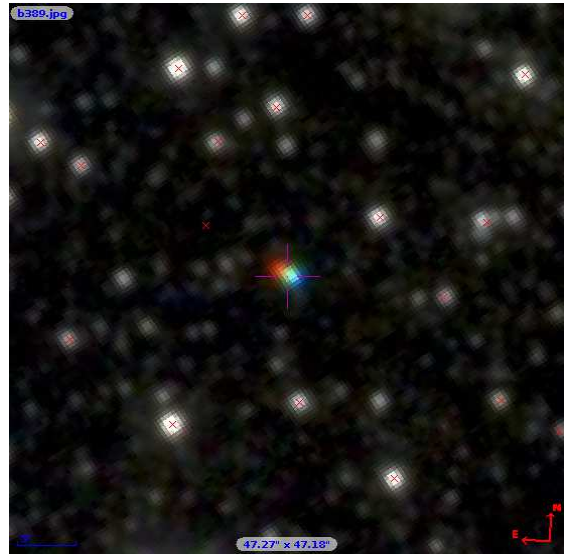}
   \caption{Finding chart for \object{VVV BD001}. Composition of three $K_{\rm s}$ band epochs (Red 2010, Green 2011, Blue 2012).
   In red crosses are objects in the 2MASS Point source catalog.}
   \label{finding}%
    \end{figure}
    
 To search for HPM objects we first made a visual inspection of VVV false color
images, where ``red'', ``green'', and ``blue'' colors correspond to three
$K_{\rm s}$-band observations taken in 2010, 2011, and 2012, respectively.
The HPM objects leave a characteristic color trace as we can
see in Fig.\,\ref{finding}

The method for finding HPM objects is as follows:
\begin{itemize}
 \item Pick three $K_{\rm s}$ band epochs (2010, 2011, 2012);
 \item Cross match the first and second epochs and then the second and third epochs, with a
 matching radius of 4'', and  $\Delta K_{\rm s}$  $\leq$ 0.3 mag.;
 \item Keep only objects with $K_{\rm s}$  $\leq$14.0 mag.;
 \item Use only objects with flags `stellar', `saturated' or `borderline stellar' \citep{2012A&A...537A.107S};
 \item Keep only objects with consistent PM $\geq$ 0.1\arcsec yr$^{-1}$ and a position angle
 consistent within 10\degr.
\end{itemize}
The WCS solution of VVV images is based on 2MASS positions, therefore the positions
and the proper motions are relative instead of absolute.
However, the bulk motion of the references stars used to derive the WCS solution is far 
below our threshold for selecting HPM stars. 
For a typical individual chip, the RMS of the WCS solution is $\sim$ 0\farcs08, 
and this is the main source of error for point sources brighter than $K_{\rm s}$  $\sim$ 14 mag. 

The candidates are confirmed via visual inspection of the false color images  
 and by blinking the VVV images, also checking the 2MASS and SuperCosmos images to see that there is an 
object in the predicted position at that particular epoch. 
Using this method we detected about 200 HPM objects at the time of writing. 
 \object{VVV BD001} is the first one among them characterized using spectroscopic 
follow-up observations.
More details on the search method, limitations, and a catalog of HPM objects
will appear in a forthcoming paper.

This object was observed with a Folded-port InfraRed
Echellette \citep[FIRE,][]{2013PASP..125..270S} at the Magellan Baade telescope,
on the night of Mar 29/30, 2013. We used the low-resolution prism
mode, with a 0.6\arcsec \,wide slit, to obtain four 63.4 sec
spectra, in ABBA nodding pattern. The usual data reduction steps
for NIR spectroscopy were followed: flat fielding, and A-B pair
subtraction to remove first-order sky emission. Then, we
traced the continua to extract one-dimensional spectra for each
of the four spectra, with the IRAF\footnote{IRAF is distributed
by the NOAO, which is operated by the AURA under cooperative
agreement with the NSF.} task {\it apall}. We removed any
remaining sky emission with the {\it apall} background removal facility task. 
Next, we applied the tracing of the objects to
subtract one-dimensional NeAr lamp spectra and wavelength-calibrated the science spectra before combining them in
wavelength space. Finally, we corrected the science spectra for
telluric absorption with observations of the A star HD 329472,
reduced following an identical procedure.

\section{Discovery and characterization of  \object{VVV BD001}}\label{analysis}
\object{VVV BD001} (\object{VVV J172640.2-273803}) is located towards the Galactic bulge, l,b= 358.85216, 
4.21662. This position makes  \object{VVV BD001} the closest BD towards the Galactic center
position and the first to be detected in this very crowded part of the sky.
This object stands out for its particular HPM and colors compatible with an L or early T dwarf 
type (see Table \ref{table1}).
We retrieved entries from the 2MASS PSC catalog \citep{2006AJ....131.1163S} and the images 
with the position expected for the object about ten years ago. 
The position and magnitudes agree with the VVV measurements. 
We also searched for optical counterparts in the SuperCosmos images, but nothing was detected or visible in any band. 
\object{VVV BD001} appeared in the DENIS I,J,$K_{\rm s}$  images, but I mag is not listed in the DENIS catalog \citep{1999A&A...349..236}.
 The object is visible in WISE images, but since it is near a brighter star, its photometry is not
in the catalog.
The Spitzer/IRAC photometry from the GLIMPSEII Legacy Survey \citep{2009PASP..121..213C} 
is available for this source and is detailed in Table \ref{table1}.
Using the expected colors for M, L, and T dwarfs from \citet{2006ApJ...651..502P}, the GLIMPSEII 
and VVV photometry suggest that the object is an early L dwarf (L0-L6)
As mentioned in Sect. \ref{Intro}, our method does not rely primarily on color selection,
but the colors (particularly those using the shorter wavelength Z and Y filters) prove useful for 
estimating spectral types for HPM objects.

   \begin{table}
   \caption{Properties of  \object{VVV BD001}.}             
   \label{table1}      
   \centering                          
   \begin{tabular}{c c}        
   \hline\hline                 
   Parameter & Value \\    
   \hline                        
            Z & 15.507 $\pm$ 0.015 mag    \\
            Y & 14.379 $\pm$ 0.012 mag            \\
            J & 13.267 $\pm$ 0.017 mag\\
            H & 12.668 $\pm$ 0.016 mag            \\
            $K_{\rm s}$ &  12.194 $\pm$ 0.015 mag \\
            IRAC 3.6$\mu$m & 11.526 $\pm$ 0.049 mag\\
            IRAC 4.5$\mu$m & 11.404 $\pm$ 0.054 mag\\
            IRAC 5.8$\mu$m & 11.178 $\pm$ 0.065 mag\\
            IRAC 8.0$\mu$m & 11.098 $\pm$ 0.047 mag\\
            Spectral Type &  $\sim$ L5 $\pm$1 \\
            $\Delta K_{\rm s}$  &  $<$ 0.05 mag \\
            $\Delta \alpha$ cos($\delta$) & -0.5455$\pm$0.004\arcsec\,yr$^{-1}$ \\
            $\Delta \delta$  & -0.3255$\pm$0.004\arcsec\,yr$^{-1}$ \\
            $\pi$ & 57$\pm$4 \,mas \\
   \hline                                   
   \end{tabular}
   \end{table}

\begin{figure}
   \centering
   \includegraphics[width=\hsize]{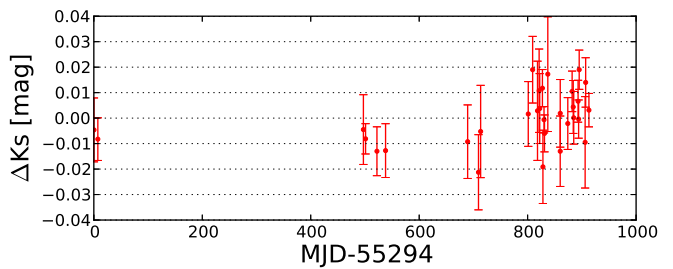}
   \caption{ Differential light curve of  \object{VVV BD001} with respect to 4 nearby stars with a 
    similar mean magnitude (|$K_{\rm s}^{BD}- K_{\rm s}^i$| $\leq$ 0.3 mag). The light curve  was
    shifted 0.015 mag, so $\langle \Delta K_{\rm s} \rangle$= 0 mag.  No evidence of periodic
    variability is detected for \object{VVV BD001}, using neither  phase dispersion minimization(PDM) nor
    Lomb-Scargle periodogram.}
    \label{variability}%
    \end{figure}

We obtained PSF photometry for every multiepoch VVV image available to date
using a new version of DoPhot \citep{1993PASP..105.1342S, 2012AJ....143...70A}.
The light curve with 31 observations of \object{VVV BD001} is shown in 
Fig. \ref{variability}, where each point of the BD light curve is the average of two observations taken 
about one minute apart. 
The same process was followed for the comparison stars. We selected the four closest stars,
with an average magnitude difference from the BD candidate that was no greater than 0.3 mag in $K_{\rm s}$ ,
and then we looked for a possible periodic signal. No signal of variability was seen for these four 
stars up to our photometric accuracy of $\sim$0.01mag.
We then averaged the magnitude of these four stars and used them to calculate the 
photometric erroproperlyr for each averaged epoch. The results can be seen in Fig. \ref{variability}

We analyzed the \object{VVV BD001} light curve using the phase dispersion minimization routine (PDM) 
\citep{1978ApJ...224..953S} and a Lomb-Scargle periodogram and found no evidence of periodic 
variability. Also from the light curve, we see no variability greater than 0.05 mag in the 
$K_{\rm s}$ band during the two years of observations with the VVV survey. 
We cannot rule out any atmosphere model using the $K_{\rm s}$  band alone, but the
VVV survey is a valuable tool for constraining long term NIR variability of BD and looking for possible
eclipses or transits.

\begin{figure}
    \centering
    \includegraphics[width=\hsize]{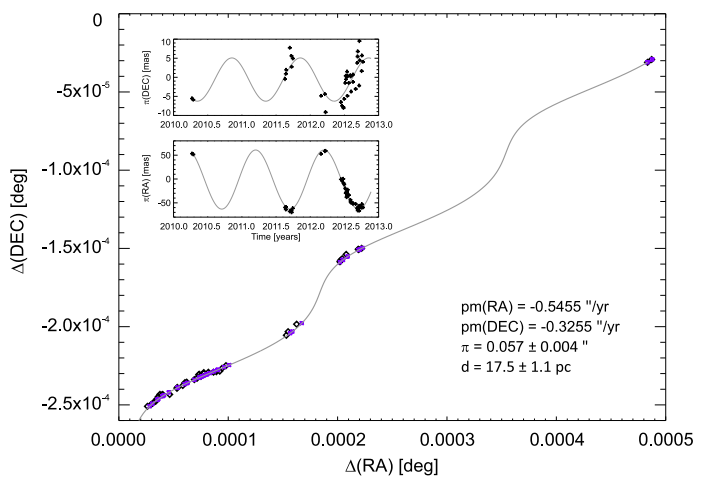}
    \caption{Proper motion and parallax movement on the sky for \object{VVV BD001}.
    Black diamonds and purple crosses give the observed and calculated best fit coordinates.
    The parallax motion in RA and DEC after removing the Proper motion are shown in the inner panel, along
    with the best fit parallax measurement.}
    \label{parallax}%
    \end{figure}

The HPM of the object suggests that it is nearby. To obtain the parallax of the target, we used 
its equatorial coordinates from CASU $ZYJH$ and all $K_{\rm s}$ epochs available to date. In total, 
there were 90 positions covering the period of March 2010 to October 2012. Five bright, 13-14 mag, and isolated 
stars without proper motion  around the target were used to obtain the corrections to a common field center 
for each epoch. Using the averaged value of the shifts, we corrected the individual epoch coordinates 
of  \object{VVV BD001}. For each epoch, the mean of the positions measured in individual frames was taken and 
its uncertainty determined from the dispersion of values in each spatial direction. In this way the 
positional errors were reduced to internal errors only ($\sim$ 7 mas). The target is presented on two 
different detectors because of the VISTA tiling procedures, and for each date there are at least 
two, and in some cases, four positions. Also $ZY$ and $JH$, together with the first epoch $K_{\rm s}$ 
images, respectively, were obtained on the same dates. Averaging date-by-date, finally, we have a 
sequence of 41 positions, used for obtaining the proper motion and the parallax. A modified version
of the {\it{Spitzer}} make\_parallax\_coords.pro: an IDL procedure designed to calculate source coordinates as seen by
an observatory was used, correcting for annual parallax and proper motion\footnote{ {\it http://irsa.ipac.caltech.edu } 
Author: S. Carey (SSC), version 12 Feb 2013}. The best parallax and proper motion fit given 
in Fig.\,\ref{parallax} leads to PM($\alpha$)=-0.5455$\pm$0.004,\arcsec yr$^{-1}$, PM($\delta$)=-0.3255$\pm$0.004,\arcsec yr$^{-1}$,
and a parallax, $\pi$=57$\pm$4\,mas which translate into a distance of d=17.5$\pm$1.1\,pc.
The tangential velocity for \object{VVV BD001} is 51.4$\pm$3.3 km s$^{-1}$, which agrees with the velocity for the blue population of L dwarfs
\citep{2009AJ....137....1F}

\begin{figure}
    \centering
    \includegraphics[width=\hsize]{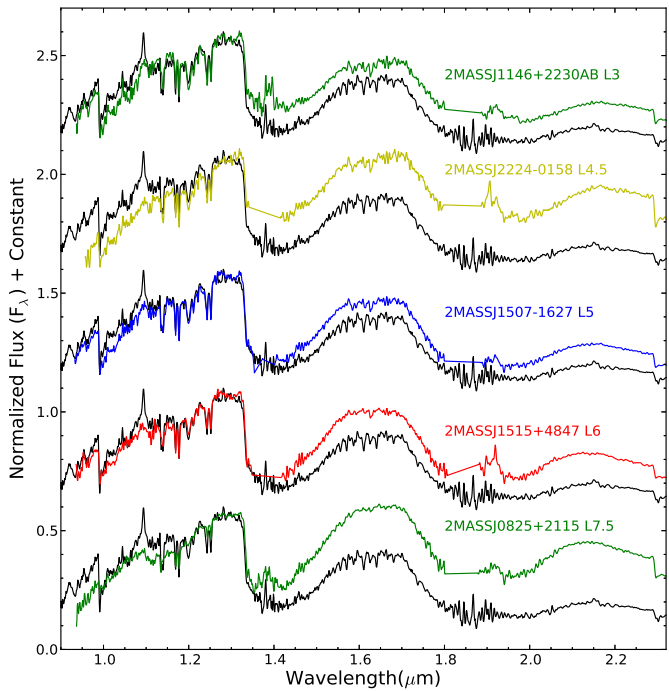}
    \caption{Spectrum of  \object{VVV BD001} (black solid line) compared with BD different spectral 
    types templates. The templates are, from top to bottom: L3 \object{2MASS J11463449+2230527}, L4.5 \object{2MASS J22244381-0158521},
    L5 \object{2MASS J15074769-1627386} L6 \object{2MASS J15150083+4847416} ; L7 \object{2MASS J08251968+2115521}. 
    The flux is normalized at 1.3 $\mu$m and each spectra has been shifted in flux for clarity.}
    \label{spectrum}
    \end{figure}

    Figure \ref{spectrum} shows the NIR spectrum of \object{VVV BD001} from 0.9-2.3 $\mu$m normalized at
    1.3 $\mu$m, in comparison with template spectra of brown dwarfs from L3 to L7
     obtained from the IRTF spectral library\footnote{{ \it{ http://irtfweb.ifa.hawaii.edu/\textasciitilde spex/IRTF\_Spectral\_Library/}}}
     \citet{2005ApJ...623.1115C}. A Gaussian kernel was used to convolve the template spectrum 
     in order to compare it with our object. 
    To classify the object, we just focused on the J band region (1.0-1.4 $\mu$m). We see in Fig.\ref{spectrum} that the 
    best match is a L5$\pm$1. An excess of flux at bluer wavelengths is clearly visible in the spectrum and was suspected 
    from the photometric data.
    Combining spectro-photometric and proper motion information, we classified  \object{VVV BD001} as an unusually blue dwarf. 
    The nature of blue L dwarfs has been discussed in \citet{2008ApJ...674..451B} and \citet{2010ApJS..190..100K} and references therein.
        
    We calculated the absolute magnitude in the 2MASS J and $K_{\rm s}$ filters, applied to the data from the 
    2MASS Point Source Catalogue, and propagated the errors in distance and photometry.
     We obtained M$_J$=12.23$\pm$0.15 mag, M$_{K_{\rm s}}$=11.06$\pm$0.15 mag. 
     After comparing M$_J$ and M$_{K_{\rm s}}$ with the expected magnitudes for a typical L5  
     \citep[see Fig.\,2 of][]{2008ApJ...672.1159B},
     our candidate is about one magnitude brighter.
     \citet{2012ApJ...760..152L} discusses the overluminosity of an unusually blue L5 dwarf
     (based on spectrophotometric distance) and favor the scenario of thin cloud condensates
     that could account for the blue nature and overluminosity.
     More follow-up observations are required to unveil the true nature of \object{VVV BD001}, 
     and given its proximity, it is a key object to observe for constraining 
     the physics behind the blue L dwarf population.

\section{Conclusions}\label{Conclusions}
   We have presented the detection of \object{VVV BD001}, the first brown dwarf detected by the VVV survey. 
   This is the first BD located towards the Galactic center,
   in the most crowded region of the sky.
   We presented new NIR photometry ranging from 0.8-2.5 $\mu$m combined with 3.6-8 $\mu$m
   available Spitzer/IRAC data. The colors, distance, and spectrum are
   compatible with an unusually blue L5$\pm$1  dwarf.
   Based on three years worth of data, we measured a total PM for \object{VVV BD001} of 0.634\arcsec yr$^{-1}$
   and a parallax of, $\pi$=57$\pm$4\,mas, yielding a  distance of 17.5$\pm$1.1 pc from the Sun.
   This makes \object{VVV BD001} a new brown dwarf, belonging to the local volume-limited sample 
   (within 20\,pc from the Sun) with well defined proper motion, distance, and luminosity.
   From our light curve we cannot exclude any atmospheric model but we provide the
   long-term behavior of the BD. Photometric multiwavelength follow-up observations on 
   shorter timescales are required to definitely determine atmospheric properties.   
   We expect to discover about two dozen BDs on the VVV survey area 
   based in the number density of brown dwarfs detected to date\footnote{the total 
   number of BDs was taken from the dwarf archive at www.dwarfarchives.org}.
   This number might be higher due to our higher sensitivity and resolution
   than previous NIR surveys.
\begin{acknowledgements}
      We gratefully acknowledge use of data from the ESO Public Survey programme ID 179.B-2002 taken with the VISTA 
      telescope, data products from  CASU, and funding from the BASAL CATA Center 
      for Astrophysics and Associated Technologies PFB-06, and the Ministry for the Economy, Development, and Tourism’s
      Programa Iniciativa Cienti\'ifica Milenio through grant P07-021-F, awarded to The Milky Way Millennium Nucleus. 
      J.C.B., D.M., M.G., R.K., J.B., acknowledges support from: PhD Fellowship from CONICYT, Project FONDECYT
      No. 1130196, the GEMINI-CONICYT Fund allocated to project 32110014. FONDECYT through grants No 1130140.
      FONDECYT through grant No 1120601, respectively.
      This publication makes use of data products from, the Two Micron All Sky Survey, which is a joint project of the University 
      of Massachusetts and the Infrared Processing and Analysis Center/California Institute of Technology, funded by 
      NASA and NSF and 
      This research made use of data obtained from the SuperCOSMOS Science Archive, prepared and hosted by the WFAU, Institute for Astronomy, University of Edinburgh, which is funded by the UK Science and Technology Facilities
      Council.
\end{acknowledgements}
\bibliographystyle{aa}
\bibliography{VVV-BD-01}
\end{document}